\documentclass[twocolumn]{aastex62}

\def\simgt{\lower.5ex\hbox{$\; \buildrel > \over \sim \;$}}
\def\simlt{\lower.5ex\hbox{$\; \buildrel < \over \sim \;$}}

\def\etal{{et~al.}}
\def\amin{\ifmmode^{\prime}\else$^{\prime}$\fi}
\def\asec{\ifmmode^{\prime\prime}\else$^{\prime\prime}$\fi}

\def\simgt{\lower.5ex\hbox{$\; \buildrel > \over \sim \;$}}
\def\simlt{\lower.5ex\hbox{$\; \buildrel < \over \sim \;$}}

\newcommand\chandra{{Chandra}}
\newcommand\xmm{{XMM-Newton}}

\newcommand\snr{G12.82$-$0.02}
\newcommand\tev{HESS~J1813$-$178}
\newcommand\tevb{HESS~J1640$-$465}
\newcommand\igr{IGR~J18135$-$1751}
\newcommand\psr{PSR~J1813$-$1749}
\newcommand\psrb{PSR~J1640$-$4631}
\newcommand\asrc{AX~J1813$-$178}
\newcommand\cl{Cl~1813$-$178}
\newcommand\tauc{$\tau_c=P/2\dot P=5600$\,yr}
\newcommand\edot{$\dot E=4\pi^2I\dot P/P^3=5.6\times10^{37}$\,erg\,s$^{-1}$}

\shortauthors{Camilo et al.}
\shorttitle{Radio Detection of \psr\ in \tev}

\begin{document}

\title{Radio Detection of \psr\ in \tev: The Most Scattered Pulsar Known} 

\author[0000-0002-1873-3718]{F.~Camilo}
\affiliation{South African Radio Astronomy Observatory, 2 Fir Street, Observatory 7925, South Africa}
\correspondingauthor{F.~Camilo}
\email{fernando@ska.ac.za}

\author[0000-0001-5799-9714]{S.~M.~Ransom}
\affiliation{National Radio Astronomy Observatory, 520 Edgemont Road, Charlottesville, VA 22903-2475, USA}

\author[0000-0003-4814-2377]{J.~P.~Halpern}
\affiliation{Department of Astronomy, Columbia University, 550 West 120th Street, New York, NY 10027-6601, USA}

\author[0000-0002-1732-5990]{D.~Anish~Roshi}
\affiliation{Arecibo Observatory, Arecibo, PR 00612, USA}
\affiliation{University of Central Florida, Orlando, FL 32816, USA}

\begin{abstract}
The 44.7\,ms X-ray pulsar in the supernova remnant \snr/\tev\ has
the second highest spin-down luminosity of known pulsars in the
Galaxy, with $\dot E=5.6\times10^{37}$\,erg\,s$^{-1}$.  Using the
Green Bank Telescope, we have detected radio pulsations from \psr\
at 4.4--10.2\,GHz.  The pulse is highly scattered, with an exponential
decay timescale $\tau$ longer than that of any other pulsar at these
frequencies.  A point source detected at this position by Dzib et
al. in several observations with the Jansky Very Large Array can
be attributed to the pulsed emission.  The steep dependence of
$\tau$ on observing frequency explains why all previous pulsation
searches at lower frequencies failed ($\tau \approx 0.25$\,s at
2\,GHz).  The large dispersion measure, $\mbox{DM}=1087$\,pc\,cm$^{-3}$,
indicates a distance of either 6.2 or 12\,kpc according to two
widely used models of the electron density distribution in the
Galaxy.  These disfavor a previously suggested association with a
young stellar cluster at the closer distance of 4.8\,kpc.  The high
X-ray measured column density of $\approx10^{23}$\,cm$^{-2}$ also
supports a large distance.  If at $d\approx12$\,kpc, \tev\ would
be one of the most luminous TeV sources in the Galaxy.

\end{abstract}

\keywords{ISM: individual objects (\tev, \snr) --- ISM: supernova remnants ---
pulsars: individual (\psr)}

\received{May 4}
\accepted{June 1, 2021}

\section{Introduction\label{sec:intro}}

\tev\ is a bright TeV source \citep{aha05,aha06} coincident with
the young shell-type radio supernova remnant (SNR) \snr\ and the
2--10\,keV X-ray source \asrc\ \citep{bro05}.  It is also detected
at 20--100\,keV as \igr\ \citep{ube05}.  \chandra\ and \xmm\ resolved
the X-ray emission into a point source and bright surrounding nebula
\citep{fun07,hel07}, evidently a pulsar and its wind nebula (PWN).
In subsequent X-ray timing observations, \citet{got09} and \citet{hal12}
discovered the $P=44.7$\,ms pulsar, with characteristic age \tauc\ and
spin-down luminosity \edot, second in power in the Galaxy only to
the Crab pulsar.

Close in projection to \psr\ is the young stellar cluster \cl\ at
a kinematic distance of 4.8\,kpc, discovered by \citet{mes08,mes11},
who proposed this as a possible birth place of the pulsar progenitor.
But \citet{hal12} argued that the discrepant measurements of optical
extinction to the cluster and X-ray absorption to the pulsar/PWN,
and X-ray absorption to a neighboring source of known distance, are
evidence that the distance to \tev\ is greater than that to \cl,
possibly as large as 12\,kpc.

Radio pulsation searches at the Green Bank Telescope (GBT) at
1.4\,GHz and 2\,GHz failed to detect a radio pulsar, with period-averaged
flux density limits of $<0.07$\,mJy and $<0.006$\,mJy, respectively
\citep{hal12}.  Despite these non-detections, a time-variable point
source at the X-ray position of \psr\ in 4.86\,GHz images from the
Very Large Array (VLA) was reported by \citet{dzi10}.  The measured
flux density was $0.18 \pm 0.02$\,mJy in 2006.  Additional Jansky
VLA (JVLA) observations by \citet{dzi18} in 2012, 2017, and 2018
continued to detect a moderately variable point source at 6\,GHz
and 10\,GHz with flux densities of $\approx0.12$\,mJy and
$\approx0.06$\,mJy, respectively.  In addition, \citet{dzi18}
performed another pulsar search at 1.4\,GHz using the 100~m Effelsberg
Telescope, obtaining an upper limit of $<0.065$\,mJy.  Since the
extrapolation of the steep spectrum of the JVLA 6\,GHz and 10\,GHz
detections to the lower frequencies of the pulsation searches greatly
exceeds the flux upper limits of those searches, it is difficult
to understand the non-detection of the pulsar and the nature of the
compact radio source.

We report pulsation searches using the GBT at frequencies of
4.4--10.2\,GHz that finally detect a highly dispersed and scattered
pulse that can be attributed to the JVLA imaged point source.  In
Section~2 we describe the observations, and explain the lack of
detection of pulsations at lower frequencies in terms of scattering.
Section~3 discusses the pulsar dispersion measure (DM) distance and
its implications for the luminosity of the TeV source, the location
of the scattering material, as well as other properties.  Suggestions
for further work on this pulsar are presented in Section~4.

\section{Observations}
\subsection{GUPPI}

\begin{figure}
\centerline{
\includegraphics[width=1.00\linewidth,angle=0,clip=true]{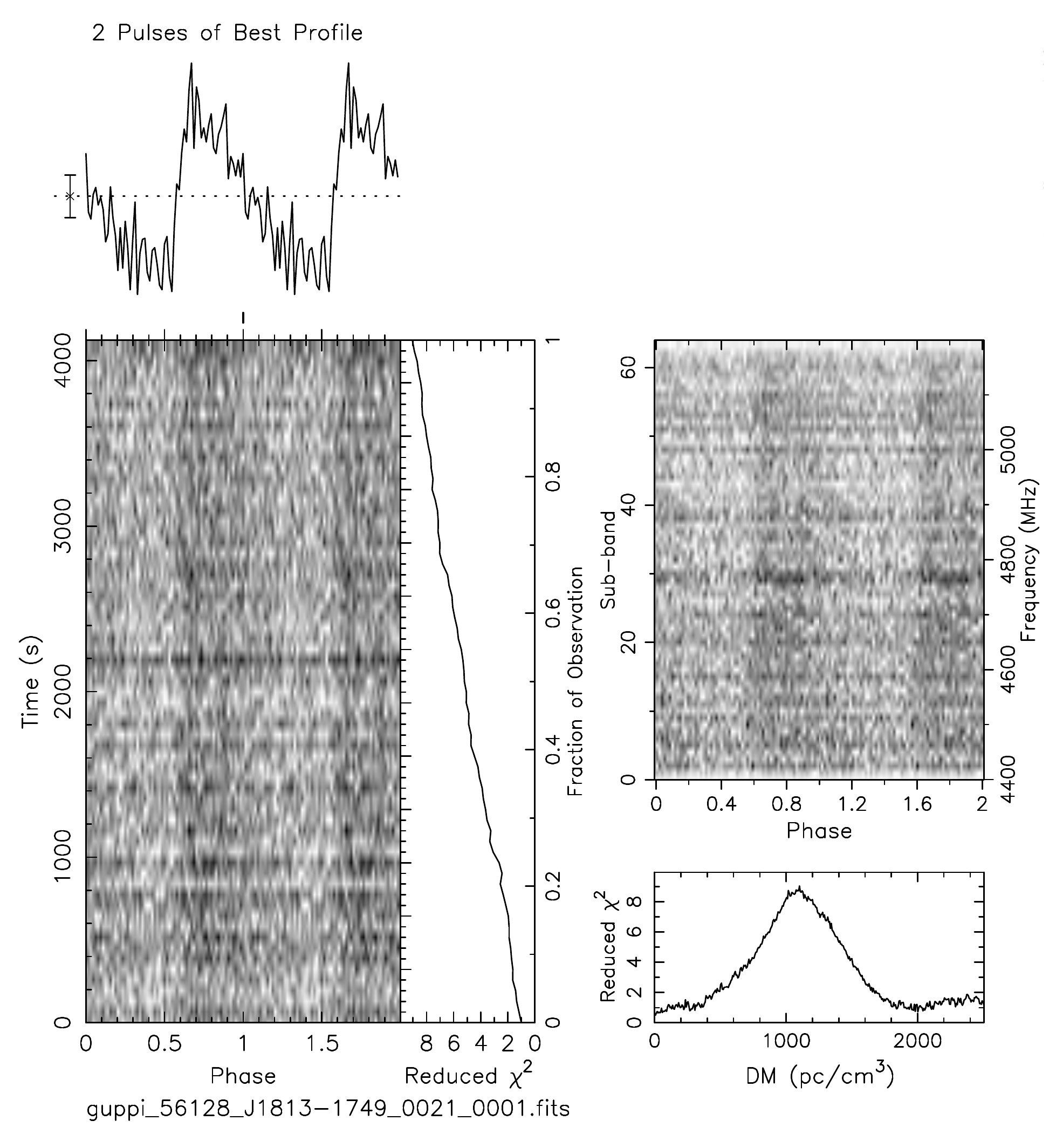}
}
\caption{Radio discovery of \psr\ on 2012 July 20 in a 70-minute
GBT observation with the old C-band receiver using GUPPI (4.4--5.2\,GHz).
The 44\,ms pulse has a clear scattering tail, which is unprecedented
among pulsars at 5\,GHz, and a nominal DM of 1102\,pc\,cm$^{-3}$.
This peak DM (based on $\chi^2$) is biased high because of scattering
at the bottom of the wide band --- the best-fit DM is 1087\,pc\,cm$^{-3}$
(see Section~\ref{sec:scatt}). The best-fit barycentric spin
period of this detection is $P=44.71259(1)$\,ms, at epoch $\mbox{MJD}
= 56128.15$.
} \label{fig:fig1} \end{figure}

On 2012 July 20 we used the GUPPI spectrometer \citep{drd+08} to
sample an 800\,MHz band centered at 4.8\,GHz, recording data from
each of 512 frequency channels every 0.163\,ms.  The integration
lasted for 70 minutes.  We used standard pulsar search techniques
implemented in PRESTO \citep{ran01}, searching in DM up to
3360\,pc\,cm$^{-3}$, twice the total Galactic value predicted for
this line of sight by the \citet{cor02} electron distribution model.
A strong detection at a barycentric period of $44.712592(12)$\,ms
was obtained, as shown in Figure~\ref{fig:fig1}, 2.3\,$\sigma$ from
an extrapolation of the incoherent X-ray ephemeris of \psr\
\citep{hal12}, which would predict $P=44.712535(22)$\,ms.  The large
DM of 1087\,pc\,cm$^{-3}$ implies a distance of $12\pm2$\,kpc
according to the \citet{cor02} model, or 6.2\,kpc in the \citet{yao17}
model.  A long, exponential scattering tail is evident even at this
high frequency, making \psr\ clearly an extremely scattered pulsar.

\begin{figure}
\centerline{
\includegraphics[width=1.00\linewidth,angle=0,clip=true]{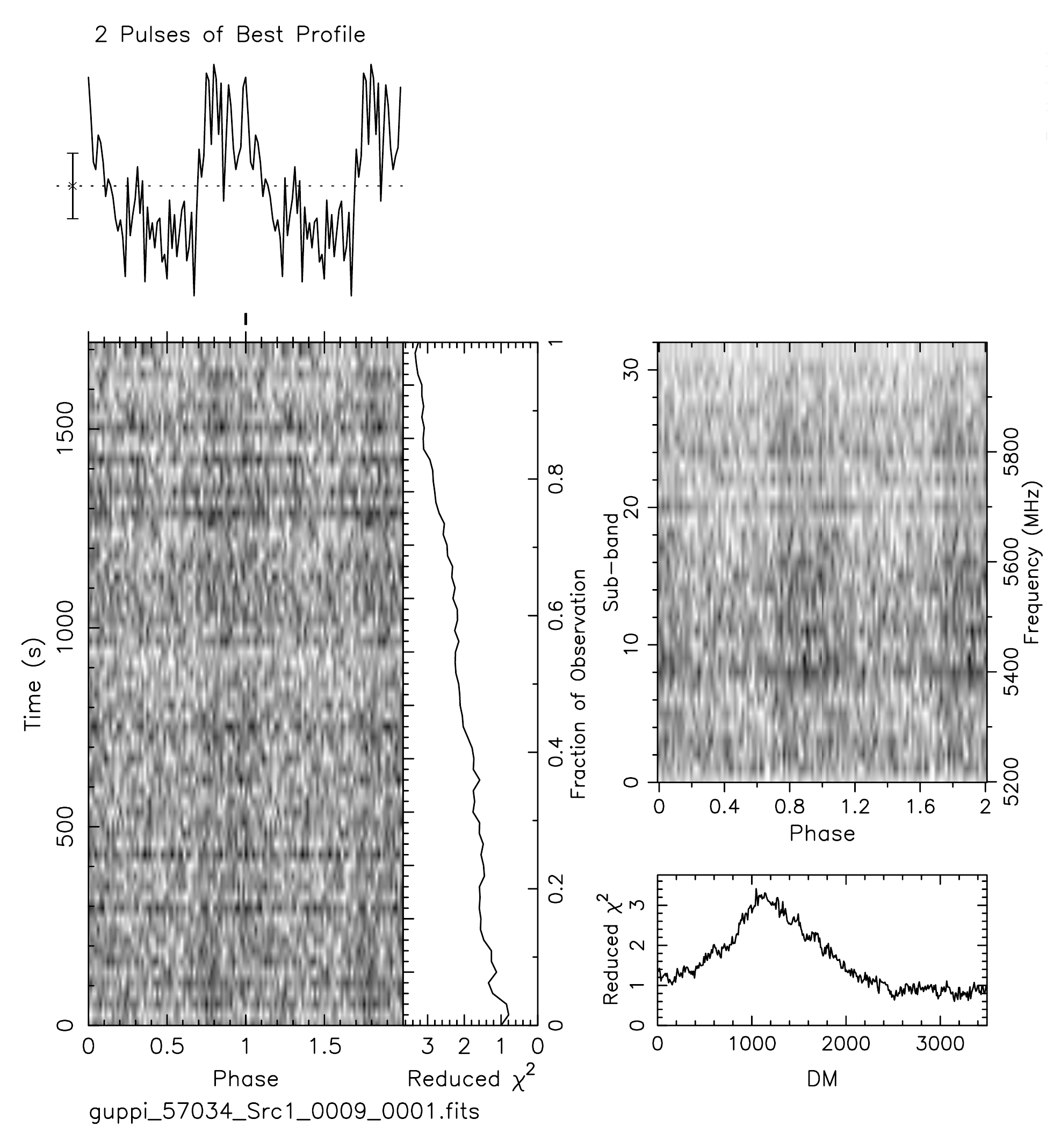}
}
\caption{\psr\ on 2015 January 12 in a 30-minute GBT observation
with the new C-band receiver using GUPPI (5.2--6.0\,GHz).
For this detection at $\mbox{MJD} = 57034.72$, $P = 44.72244(4)$\,ms.
}
\label{fig:fig2}
\end{figure}

We made another observation of \psr\ on 2015 January 12, in the
frequency range 5.2--6.0\,GHz, this time using a new C-band receiver.
The pulsar was clearly detected in the 30 minute observation
(Figure~\ref{fig:fig2}) with a period of $44.722438(35)$\,ms, this
time with a somewhat narrower profile consistent with expectation
from the typical $\nu^{-4}$ dependence of the scattering timescale
on observing frequency $\nu$.  A formal measurement of the scattering
timescale is made in Section~\ref{sec:scatt}.  The period derivative
between the two radio detections is $\dot P = 1.2570(37)\times10^{-13}$,
very close to the earlier X-ray measured value of
$1.26545(64)\times10^{-13}$ \citep{hal12}.

\subsection{VEGAS\label{sec:VEGAS}}

\begin{figure}
\centerline{
\includegraphics[width=1.00\linewidth,angle=0,clip=true]{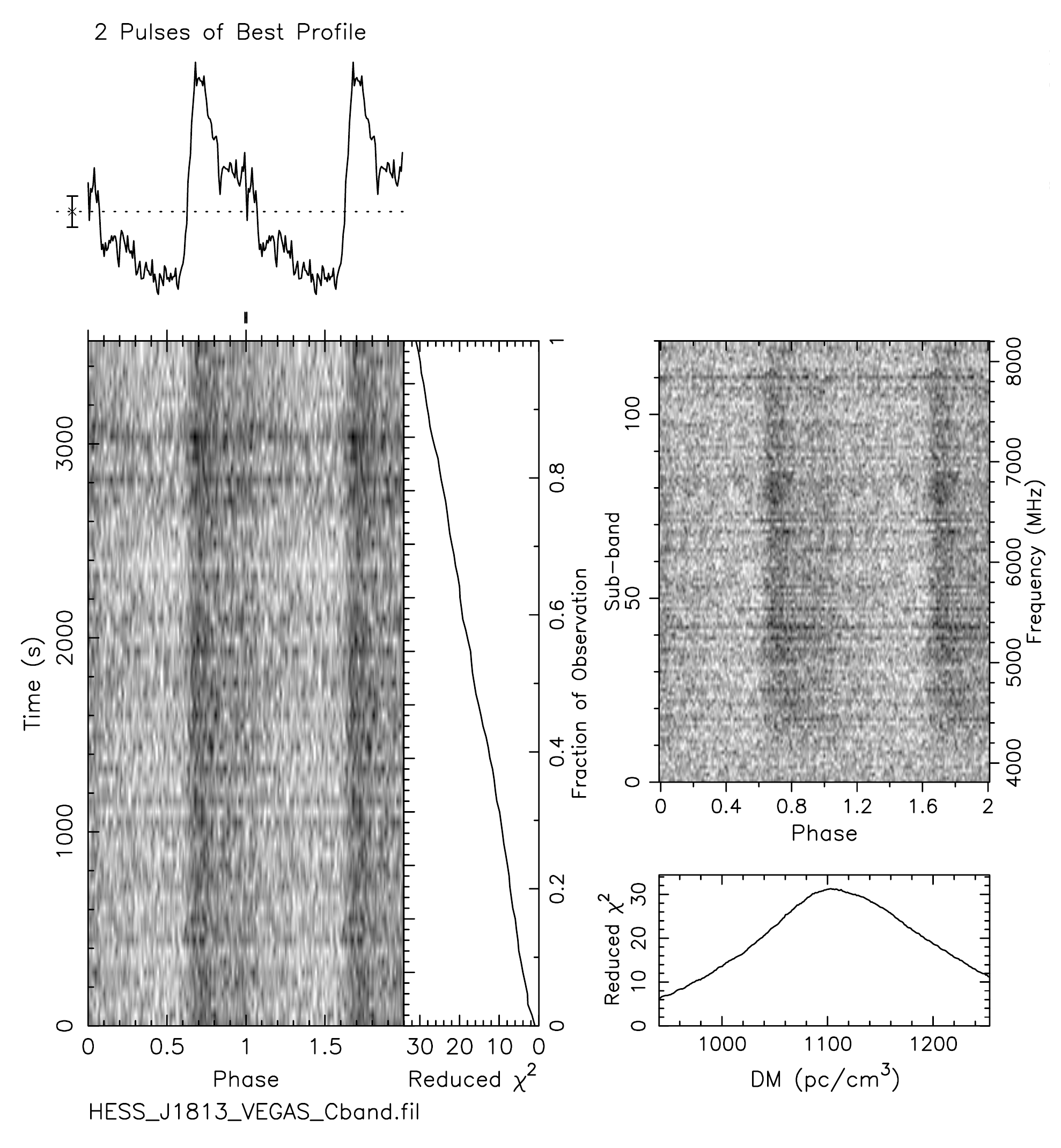}
}
\caption{\psr\ on 2015 January 10 in a 60-minute GBT observation
with the new VEGAS spectrometer in C-band (3.8--8.2\,GHz).
For this detection at $\mbox{MJD} = 57032.57$, $P = 44.722434(8)$\,ms.
}
\label{fig:fig3}
\end{figure}

\begin{figure}
\centerline{
\includegraphics[width=1.00\linewidth,angle=0,clip=true]{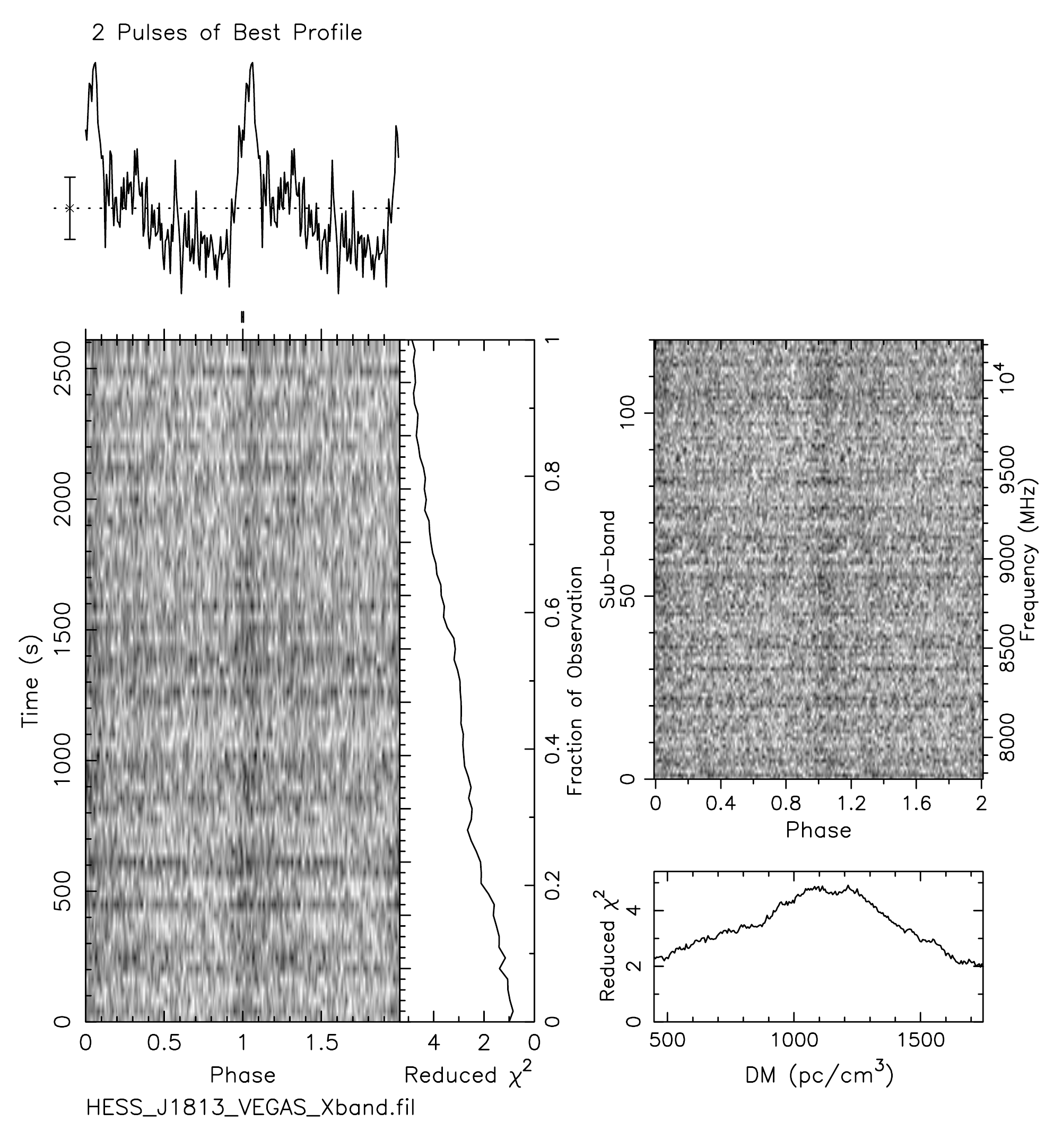}
}
\caption{\psr\ on 2015 January 10 in a 45-minute GBT observation
with VEGAS in X-band (7.8--10.2\,GHz).
For this detection at $\mbox{MJD} = 57032.62$, $P = 44.72239(3)$\,ms.
}
\label{fig:fig4}
\end{figure}

On 2015 January 10 we also observed \psr\ with the new VEGAS
spectrometer \citep{pbb+15}, using 0.5\,ms sampling.  A 60 minute
observation in C-band (3.8--8.2\,GHz) and a 45 minute observation
in X-band (7.8--10.2\,GHz) each detected the pulsar (Figures~\ref{fig:fig3}
and \ref{fig:fig4}).  The broader bandpass of VEGAS provides a
higher signal-to-noise ratio (S/N) pulse profile in C-band, and
a weaker pulse at X-band, with no scattering evident.  These
higher-quality data show that the pulse shape is complex.  Rather
than a single narrow pulse with an exponential tail, there appear
to be two components intrinsic to the pulse profile.

\begin{figure}
\centerline{
\includegraphics[width=1.00\linewidth,angle=0,clip=true]{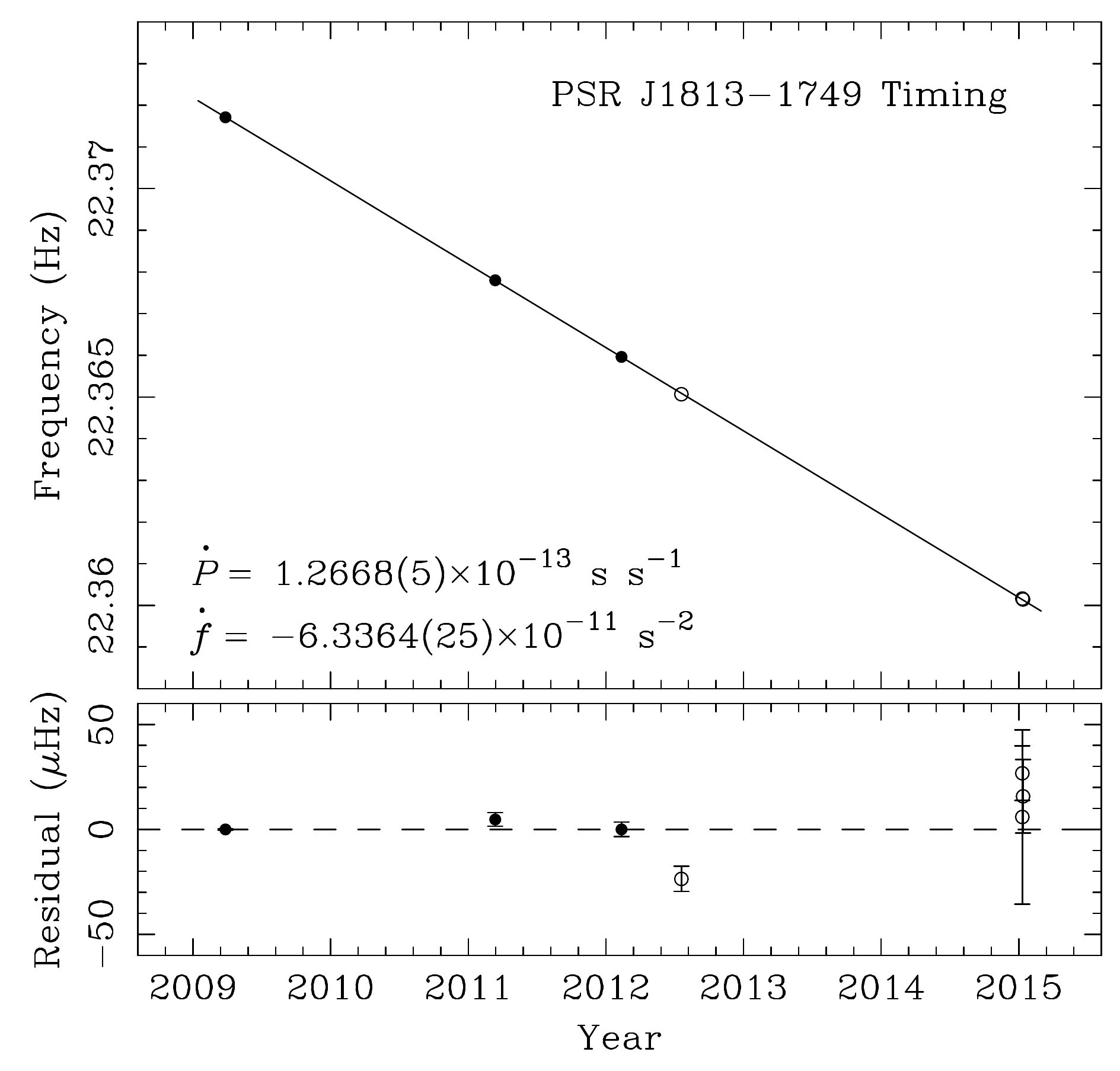}
}
\caption{Timing of \psr.  Filled circles are X-ray measurements
from \citet{hal12}; open circles are radio measurements reported
here.
}
\label{fig:fig5}
\end{figure}

A fit to all the radio and prior X-ray measured periods is shown
in Figure~\ref{fig:fig5}, which yields an overall $\dot P =
1.2668(5)\times10^{-13}$.  The fit is not perfect, however, likely
because of intervening glitches and/or timing noise. This $\dot P$
is within 3.2\,$\sigma$ of a longer-term average value based only
on X-ray measurements \citep{ho20}.

\subsection{Pulse Broadening of \psr\ \label{sec:scatt}}

\begin{figure}
\centerline{
\includegraphics[width=1.00\linewidth,angle=0,clip=true]{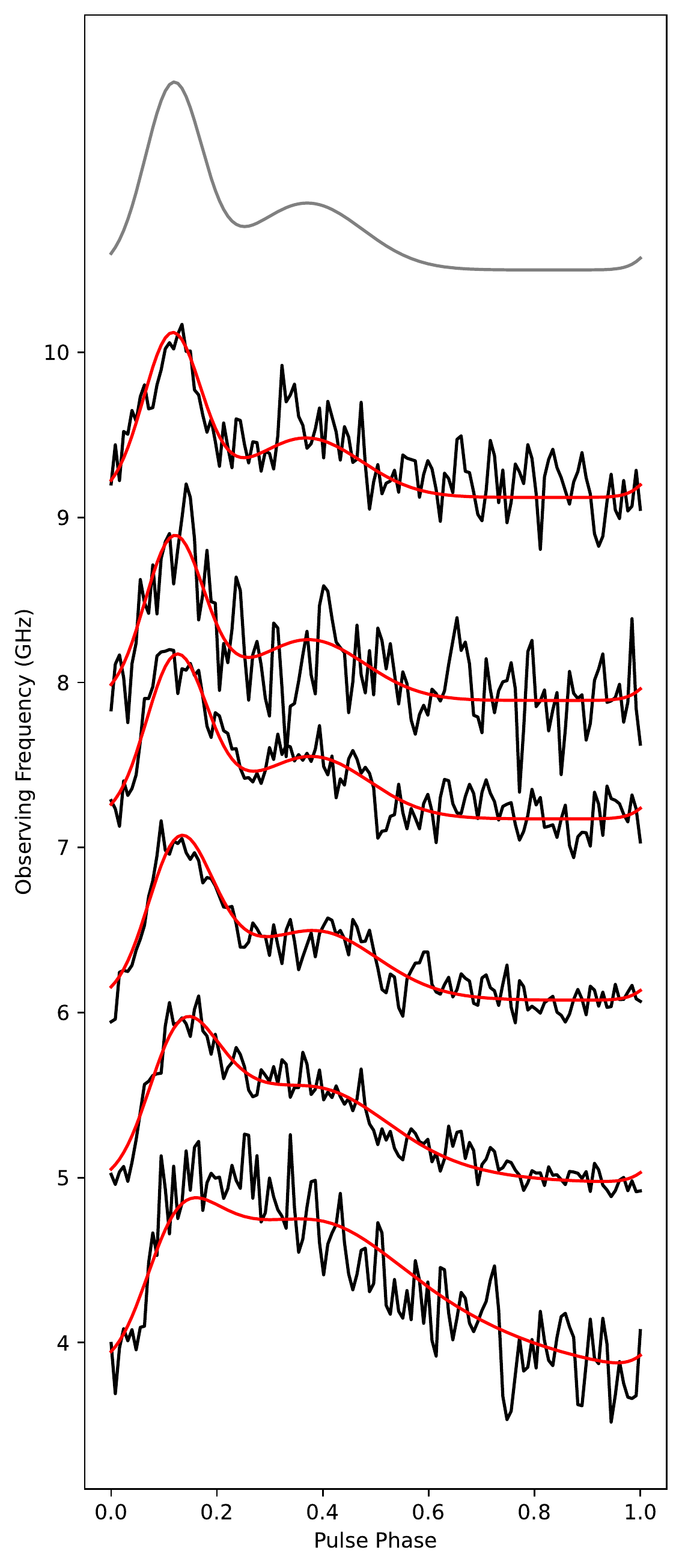}
}
\caption{Scattering fit to \psr\ profiles at C- and X-bands.  The
lower four black traces are the observed profiles in four C-band
subbands (cf., Figure~\ref{fig:fig3}), and the upper two black
traces are two X-band subband profiles (cf., Figure~\ref{fig:fig4}).
The gray trace is the assumed infinite-frequency intrinsic pulse
profile based on a fit to the X-band data (Figure~\ref{fig:fig4}).
The red traces are model profile fits of the six subbands (see
Section~\ref{sec:scatt} for details).
}
\label{fig:fig6}
\end{figure}

In order to estimate the amount of pulse broadening for \psr, we
performed a simultaneous fit to the C-band and X-band VEGAS profiles
(Figures~\ref{fig:fig3} and \ref{fig:fig4}) using the {\tt emcee}
package for MCMC analyses \citep{for13}. We assumed a simple one-sided
exponential pulse broadening model, with scattering timescale
$\tau(\nu)$ proportional to $\nu^{-4}$ \citep{osw21}.  Since
we do not know the intrinsic pulse shape in these observing bands,
we fitted a two-gaussian model to the X-band data (see the top gray
line in Figure~\ref{fig:fig6}), and assumed that this is the infinite
frequency pulse profile.

We split the C-band data into four equal subbands as a function of
$\nu$, and the X-band data into two equal subbands (black traces
in Figure~\ref{fig:fig6}). Splitting into subbands is necessary
because the timescale of scattering within the broad bands varies
considerably from bottom to top as a result of the strong $\nu^{-4}$
dependence of $\tau$. The MCMC fits included multiple nuisance
parameters since the correct relative scalings between any of the
six fitted profiles, the absolute alignment between the two different
observations, and the true DM of the pulsar (the measured value is
slightly biased by scattering) are all unknown.

The model-dependent estimate of the pulse broadening timescale at
1\,GHz is $\tau_{\rm 1\,GHz} = 4.14\pm0.11$\,s, where the quoted
error is purely statistical. Systematics resulting from the low-S/N
subband detections, the unknown intrinsic profile shape, and the
simple scattering model could substantially modify that value.  In
any case, this scattering timescale (equivalent to 1.1\,s at 1.4\,GHz
and 0.26\,s at 2\,GHz) would smear out almost all of the pulsed
flux for this 44\,ms pulsar at lower frequencies, accounting for
its non-detection in pulsed searches at 1.4\,GHz and 2\,GHz (see
Section~\ref{sec:intro}). Even at 4.4\,GHz (the bottom of the
discovery radio band, cf. Figure~\ref{fig:fig1}), the implied 11\,ms
scattering timescale is 25\% of the rotation period.

The MCMC-fit value of the dispersion measure is $\mbox{DM} = 1087
\pm 0.5$\,pc\,cm$^{-3}$, where the quoted error is purely statistical.
This is model dependent (e.g., it assumes that there is no intrinsic
frequency evolution of the pulse profile, and that
$\tau(\nu)\propto\nu^{-4}$) and there is some covariance with other
fit parameters. We estimate a systematic DM error of $\approx
3$\,pc\,cm$^{-3}$.

\subsection{Pulsed Flux Density of \psr\ \label{sec:flux}}

We obtain the period-averaged flux density $S_\nu$ for \psr\ at
C-band and X-band by applying the standard modified radiometer
equation to the VEGAS detections (Section~\ref{sec:VEGAS}).

Prior to doing the VEGAS pulsar observations, we optimized the
pointing and focusing of the subreflector/receiver system. The
on-line calibration procedure for these observations returned a
system temperature of 18.8 and 29.7\,K for C-band and X-band,
respectively. In turn, the nominal GBT gain for these bands is 1.87
and 1.8\,K\,Jy$^{-1}$.

Using the above in the modified radiometer equation with parameters
appropriate for the VEGAS observations (Figures~\ref{fig:fig3} and
\ref{fig:fig4}) yields $S_{\rm 6\,GHz} \approx 0.06$\,mJy and
$S_{\rm 9\,GHz} \approx 0.06$\,mJy. These are crude estimates; e.g.,
the C-band value may be biased low by scattering, and the
X-band detection suffers from low S/N.

\section{Discussion}
\subsection{Scattering Timescale and JVLA Point Source}

\psr\ doubtless accounts for the JVLA point source with $S_{\rm
6\,GHz} \approx 0.12$\,mJy and $S_{\rm 10\,GHz} \approx 0.06$\,mJy
\citep{dzi18}. Firstly, these flux densities are comparable to those
estimated from the pulsed detections (Section~\ref{sec:flux}).

Also, despite the extreme pulse scattering, angular broadening is
not expected to be detectable in the existing images.  A scattering
timescale $\tau$ corresponds to angular broadening of the image
with FWHM $$\theta=\sqrt{{8\,{\rm ln}\,2\,c\,(d-s)\,\tau \over
d\,s}}\eqno{(1)}$$ for a source at a distance $d$ scattered by a
thin screen at a distance $s$ \citep{cor97}.  Our fitted scattering
timescale at 6\,GHz is $\tau_{\rm 6\,GHz}=3.2$\,ms.  Assuming
$d=6.2$\,kpc, and $s = d/2$, the predicted angular broadening is
$\theta_{\rm 6\,GHz}\approx0.\!^{\prime\prime}034$, which is much
smaller than the typical $1.\!^{\prime\prime}6\times0.\!^{\prime\prime}8$
beam size of the \citet{dzi18} JVLA B-array observations, consistent
with their detection of the pulsar as an unresolved source.

\subsection{Distance, Associations, and Energetics}

Arguments pertaining to the distance and possible associations with
\psr/\snr\ were detailed by \citet{hal12}, and are summarized here.
The young stellar cluster \cl\ is centered $4.\!^{\prime}4$ southwest
of \snr.  \citet{mes11} concluded that \cl\ is one of several
clusters belonging to the massive star forming region W33 at
$(\ell,b)=(12\fdg8,-0\fdg2)$, and determined that it has an age of
4--4.5\,Myr, which is ideal for the production of neutron stars.
They derived a spectrophotometric distance to \cl\ of $3.6 \pm
0.7$\,kpc, and a kinematic distance of $4.8\pm 0.3$\,kpc from the
radial velocity of the brightest star in the cluster.  At 4.8\,kpc,
the $2.\!^{\prime}5$ diameter of the SNR corresponds to a radius
of 1.7\,pc.  Subsequent to the above work, \citet{imm13} measured
a distance of $2.4^{+0.17}_{-0.15}$~kpc to the main W33 complexes
using trigonometric parallaxes of water masers, in contradiction
to a previously assumed kinematic distance of 3.7\,kpc, which
suggests that \cl\ is not associated with W33.

In any case, the absorbing column densities to \psr\ and \cl\ are
discrepant. The X-ray measured $N_{\rm H} = (10\pm1) \times
10^{22}$\,cm$^{-2}$ to \snr\ \citep{hel07} is very high.  An even
larger value of $N_{\rm H} = (13.1\pm 0.9) \times 10^{22}$\,cm$^{-2}$
was derived by \citet{ho20} analyzing the same and newer data.  In
comparison, the average visual extinction to \cl\ of $A_V = 9.1$
\citep{mes11} corresponds to an equivalent X-ray absorption of
$N_{\rm H} = 2 \times 10^{22}$\,cm$^{-2}$ according to the relation
$N_{\rm H} = 2.21 \times 10^{21}A_V$\,cm$^{-2}$ \citep{guv09}.  The
highest extinction for a cluster member is $A_V = 17$ (equivalent
to $N_{\rm H} = 3.8\times 10^{22}$\,cm$^{-2}$), which still does
not come close to matching the X-ray $N_{\rm H}$.  Taking into
account the observed column density of molecular gas together with
its velocity information from CO, it appears that the X-ray absorption
is consistent with any distance in the range 5--12\,kpc.  The wide
possible range is due to the uncertain partition of molecular gas
between the near and far branches of the double-valued rotation
curve.   The newly measured DM is consistent with such distances
as it predicts $d=12\pm2$\,kpc according to the \citet{cor02}
electron distribution model or 6.2\,kpc in the \citet{yao17} model.

Further evidence for a large distance comes from a comparison with
the X-ray absorption to the bright LMXB GX~13+1 that lies only
$0\fdg7$  from \snr\ along the Galactic plane.  The distance to
GX~13+1 was estimated as $7\pm1$\,kpc \citep{ban99} from the
spectroscopic classification (K5III) of its companion star and
extinction variously estimated as $A_V=13.2$--17.6, while its X-ray
column density is less than one-third that to \psr.  This suggests
that the latter is farther than 7\,kpc, compatible with the DM
derived values.  The \ion{H}{2} regions of the intervening W33
complex may, however, contribute to the scattering of \psr, similar
to the conclusion of \citet{dex17} concerning other pulsars and
\ion{H}{2} regions in the inner Galaxy.

If \tev\ is located at $d=12$\,kpc, its $>200$\,GeV luminosity would
be $\approx3\times 10^{35}$\,erg\,s$^{-1}$, which is $<1\%$ of the
$\dot E$ of \psr.  It and \tevb\ associated with \psrb\ \citep{got14},
also at a distance of 12\,kpc, would have nearly the same luminosity
and would be two of the most powerful TeV sources in the Galaxy.

\subsection{Comparison with Other Pulsars\label{sec:comparison}}

\begin{figure}
\centerline{
\includegraphics[width=1.00\linewidth,angle=0,clip=true]{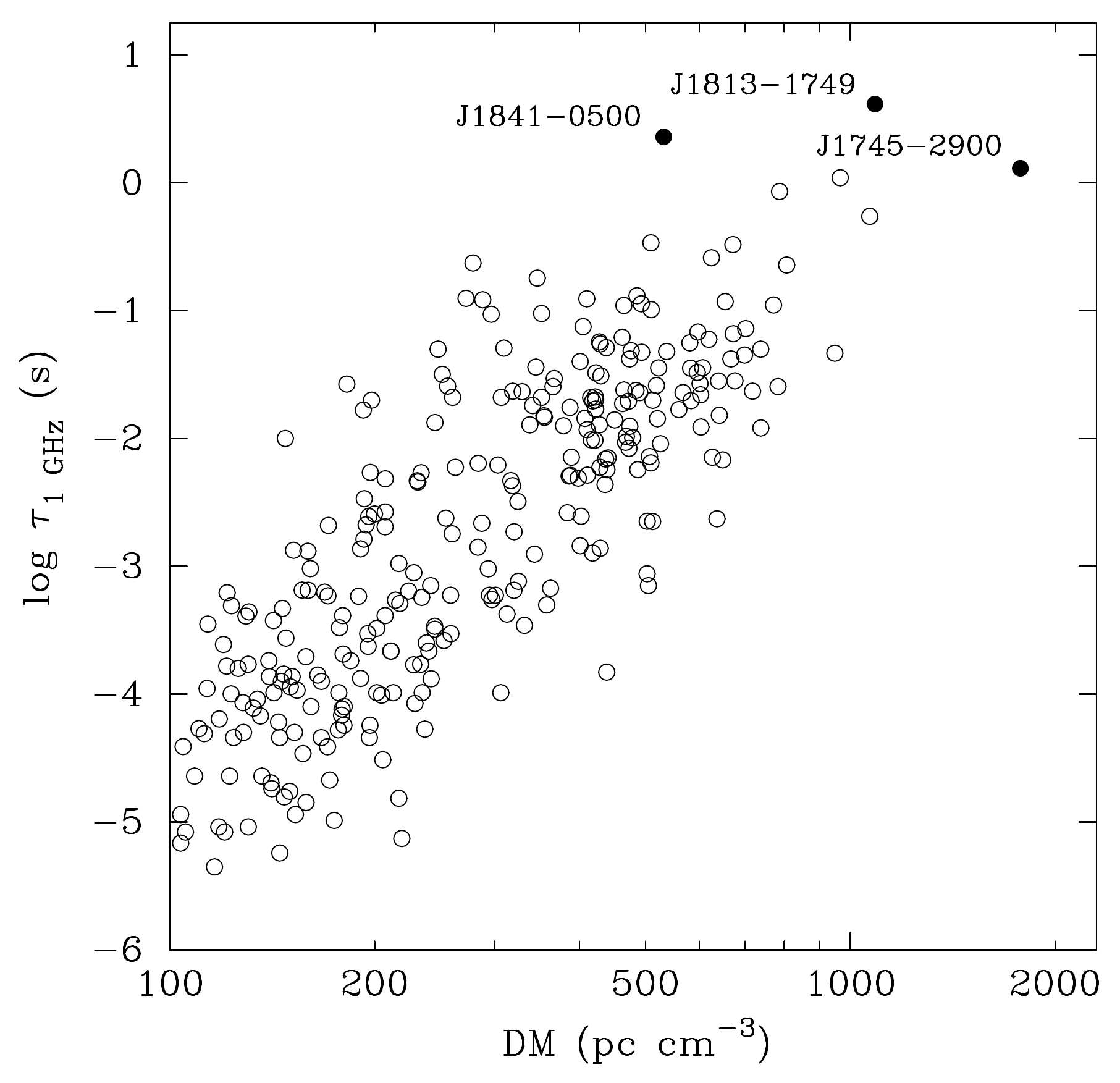}
}
\caption{Scattering timescale at 1\,GHz versus dispersion measure
for pulsars with the highest DM.  Open circles are from the ATNF
Pulsar Catalog \citep{man05}, supplemented with recent measurements
from MeerKAT \citep{osw21}. Labelled pulsars (filled circles) are
discussed in Section~\ref{sec:comparison}.
}
\label{fig:fig7}
\end{figure}

Figure~\ref{fig:fig7} graphs scattering timescale against dispersion
measure for the highest DM pulsars.  Data are from the ATNF Pulsar
Catalog\footnote{https://www.atnf.csiro.au/research/pulsar/psrcat/}
\citep{man05} version 1.64, supplemented with recent measurements
from MeerKAT \citep{osw21}.  Although there is a general correlation
of $\tau_{\rm 1\,GHz}$ with DM among pulsars, the spread in $\tau_{\rm
1\,GHz}$ at a given DM is several orders of magnitude \citep{loh01,lew15}.
This could be due to the placement of the scattering material or
inadequacy of the isotropic thin-screen model.  The scattering
timescale of \psr\ makes it the most scattered pulsar known,
although it falls within the spread of the correlation with DM.
It is not more of an outlier in Figure~\ref{fig:fig7} than, e.g.,
PSR~J1841$-$0500, the intermittent pulsar discovered by \citet{cam12}.

The Galactic Center (GC) magnetar PSR~J1745$-$2900 has
$\mbox{DM}=1778$\,pc\,cm$^{-3}$, the highest of any known radio
pulsar, but its scattering timescales of $\tau_{\rm
4.8\,GHz}=3.3\pm0.6$\,ms and $\tau_{\,\rm 1\,GHz}=1.3$\,s \citep{spi14}
are smaller than those of \psr.  \citet{bow14} used the equality
of the image broadening of PSR~J1745$-$2900 and Sgr A$^*$ to argue
that they are scattered by the same screen, which Equation (1)
places at 2--3\,kpc from Earth rather than near the GC.  Such a
determination of the scattering location cannot yet be made for
\psr\ because its image size has not been measured.

What else we know from pulsars close to the line of sight is
summarized in Table~\ref{tab:pulsars}, which lists data on four
pulsars at $<0\fdg5$ angular distance from \psr.  All four have DM
distances greater than the parallax distance of W33, but none is
as distant as \psr.  The two with the largest DM (other than \psr)
are closest to the Galactic plane, but do not have scattering
timescales measured.  The two with reported scattering measurements
have timescales much smaller than that of \psr.  Together, these
data argue that \psr\ lies behind additional, distant scattering
material that is not associated with the W33 complex and not traversed
by the other pulsars' sightlines.

\begin{deluxetable}{lrrlccc}
\tablewidth{0pt}
\tablecolumns{7}
\tablecaption{Pulsars within $0\fdg5$ of \psr}
\tablehead{
	\colhead{Name} & \colhead{$\ell$} & \colhead{$b$} & \colhead{DM\tablenotemark{a}} &
	\colhead{$d$\tablenotemark{b}} & \colhead{$\tau_{\,\rm 1\,GHz}$\tablenotemark{a}} & \colhead{$\theta$} \\
	\colhead{(PSR)} & \colhead{($^{\circ}$)} & \colhead{($^{\circ}$)} &
	\colhead{(pc\,cm$^{-3}$)} & \colhead{(kpc)} & \colhead{(s)} &
	\colhead{($^{\circ}$)}
}
\startdata
J1811$-$1736 & 12.82 & $0.43$  & 473.93(4) & 4.4 & 0.042(1) & 0.45 \\
J1812$-$1733 & 12.90 & $0.38$  & 509.8(1)  & 4.5 & 0.102(1) & 0.41 \\
J1813$-$1749 & 12.81 & $-0.02$ & 1087(3)   & 6.2 & 4.14(11) & \dots \\
J1814$-$1744 & 13.02 & $-0.21$ & 792(16)   & 5.0 & \dots    & 0.28 \\
J1815$-$1738 & 13.17 & $-0.27$ & 724.6(2)  & 4.9 & \dots    & 0.44 \\
\enddata
\tablenotetext{a}{Uncertainties on the last digits are in parentheses.}
\tablenotetext{b}{DM distance from the \cite{yao17} model.}
\label{tab:pulsars}
\end{deluxetable}

\subsection{Proper Motion and Age}

Recently, \citet{ho20} reported a proper motion of
$0.0655\pm0.0114$\,arcsec\,yr$^{-1}$ for \psr\ using three \chandra\
images over 10 years.  This implies a high tangential velocity of
$v_t=1490$\,km\,s$^{-1}$ at the 4.8\,kpc distance assumed in previous
studies, but an even larger $v_t=1925$\,km\,s$^{-1}$ if it is at
the \citet{yao17} distance of 6.2\,kpc.  This would be larger than
any well-measured velocity for a neutron star (cf., \citealt{del19}),
a result that invites skepticism as well as exploration of its
implications.

For one, \psr\ is only $\approx20^{\prime\prime}$ from the center
of the radio shell of \snr, unlike other high-velocity pulsars that
have either escaped their shells or show other morphological evidence
of high velocity in the structure of their PWNe (for a review, see
\citealt{kar17}).  These properties might, however, be reconciled
if \snr\ has a very young age of $\approx300$\,yr.  This would
possibly make \psr\ the youngest known neutron star in the Galaxy.
If so, it would also be interesting if such a young SNR is accompanied
by dense plasma that could cause the extreme scattering of the radio
pulses.   Note from Equation (1) that for a scattering screen very
close to the pulsar ($s\approx d$), a given image broadening would
correspond to a very large scattering timescale $\tau$.

On the other hand, countervailing evidence against such a young age
is found in the broadband spectral energy distribution of \tev,
which is more like those of evolved SNRs, and has been modelled
most recently with an age of 2500\,yr \citep{zhu18}.  In summary,
there would be obstacles to attributing the scattering of \psr\ to
a very young age.

\section{Conclusions and Suggestions for Further Work} 

Pulsation searches at high radio frequency were necessary to overcome
the extremely long scattering timescale of \psr, the largest among
known pulsars.  The GBT pulsed radio detection at 4.4--10.2\,GHz
with $\mbox{DM}=1087$\,pc\,cm$^{-3}$ provides additional evidence
of a larger distance than once assumed, now favoring $d\approx6$--14\,kpc.
If at the higher end of this range, \tev\ may be one of the
most luminous TeV source in the Galaxy, at the expense of $<1\%$
of the spin-down power of \psr.  A complete census of pulsars in
the Galaxy, even energetic ones such as \psr, remains a challenging
prospect if there are more lines of sight with such long scattering
timescales.

Previous modelling of the spectral energy distribution of \tev\ as
leptonic emission from an evolving PWN in an expanding SNR
\citep{fan10,zhu18} used a distance of 4.7\,kpc.  Models should
also be recalculated for the now favored larger distances.  However,
it is not universally accepted that the PWN is the source of the
TeV emission, as opposed to the SNR shell or hadronic interactions
with the environment \citep{tor14}.

Since the DM distance still has a large uncertainty, it would be
worthwhile to obtain a more sensitive observation of 21~cm \ion{H}{1}
absorption against \snr\ to pin down a kinematic distance. Future
\chandra\ observations are needed to confirm or disprove the reported
high proper motion of \psr, which would imply a very young age for
the system, and possibly be related to the extreme pulse scattering
timescale.  VLBI at 5\,GHz could also measure proper motion, as
well as image broadening that could be used to deduce the location
of the scattering medium.

Finally, the detection in Figure~\ref{fig:fig3} indicates that the
combination of C-band and VEGAS can produce high-S/N pulse
profiles in only a few minutes of observing time, opening opportunities
to obtain a phase-connected timing solution for \psr.  This might
lead to the determination of its spin-down braking index, which is
an important parameter to model the broadband (spectral) evolution
and energetics of its PWN and \tev.

\acknowledgements

This material is based upon work supported by the Green Bank
Observatory which is a major facility funded by the National Science
Foundation operated by Associated Universities, Inc.  We thank Ryan
Lynch for assistance with VEGAS data analysis. The National Radio
Astronomy Observatory is a facility of the National Science Foundation
operated under cooperative agreement by Associated Universities,
Inc. SMR is a CIFAR Fellow and is supported by the NSF Physics
Frontiers Center award 1430284. We are grateful to Michael Kramer
and Ralph Eatough for obtaining trial observations at Effelsberg.

\facility{GBT}

\end{document}